\documentclass[12pt, draftclsnofoot, onecolumn]{IEEEtran}
\ifCLASSOPTIONcompsoc
  \usepackage[nocompress]{cite}
\else
  \usepackage{cite}
\fi
\usepackage{amsmath}
\usepackage[ruled,vlined,linesnumbered]{algorithm2e}
\usepackage{array}
\usepackage{float}
\usepackage{graphicx}
\usepackage{pgfplots}
\pgfplotsset{compat=newest}
\usepackage{smartdiagram}
\usepackage{enumitem}
\newlist{steps}{enumerate}{1}
\setlist[steps, 1]{label = Step \arabic*:}

\begin{document}
\title{Authenticated Hand-Over Algorithm\\ for Group Communication}

\author {Y\"{u}cel~Ayd{\i}n,
       ~G\"{u}ne\c{s}~Karabulut~Kurt~\IEEEmembership{Senior Member} and {Enver~Ozdemir~\IEEEmembership{Member} }  

}

\maketitle
\begin{abstract}
Shamir or Blakley secret sharing schemes are used for the authentication process in the studies before, but still secure group authentication and hand-over process remain as challenges in group authentication approaches. In this study, a novel method is proposed to provide a secure group authentication. The proposed approach also enables a hand-over process between groups by using Lagrange's polynomial interpolation and Weil pairing in elliptic curve groups for wireless networks with mobility support. One of the advantages of our proposed scheme is that the computational load for a member in the group is lower than the other schemes in the state-of-the-art. It is also possible to authorize many users at the same time, not one-to-one as in the group authentication methods in current cellular networks including Long Term Evolution (LTE). Another advantage that is not covered in other secret sharing methods is that the proposed approach constitutes a practical solution for the hand-over of members between different groups. We have also proposed a solution for replay and man-in-the-middle attacks in secret exchange.
\end{abstract}


\IEEEpeerreviewmaketitle

\section{Introduction}
\IEEEPARstart{A}{uthentication} is a process for ascertaining an entity really is who it claims to be \cite{1}.  It is one of the most important processes in access control chain as all other security and data transmission operations will follow after the authentication process.  There will be many interconnections in a wirelessly connected and distributed environment in the future. In such an environment, the authentication of those who have come together to form a group is not as an easy task. Additionally, the elements that perform this authentication process have very little resources and computational power. From security standpoint, the worst case is that all the devices in the communication network are mobile and this case will be a reality in the near future, that is, the number of mobile devices and the distributed networks will increase dramatically.
\\
\indent In a highly distributed environment, the users create groups within themselves according to certain characteristics or their coverage area. Due to the difficulty of making one-to-one authentication among all the members within the group, the concept of group authentication paradigm has recently emerged. After the group authentication process, the next step is to authenticate  members from other groups, which is called hand-over between groups. The research has been mostly focused  on group authentication in a wireless environment, but the hand-over problem has not yet been addressed before. Traditional authentication process includes one claimer that requests authentication and one prover that approves the claims. This process can be called one-to-one authentication. One-to-one authentication is no longer applicable in a distributed environment. If $n$ users want to authenticate each other, one user should repeat the authentication process $n-1$ times and this requires
approximately $O(n^2)$ communications. Many-to-many authentication, which is referred to as group authentication is the new scheme for the complex, mobile and crowded networks. The main idea of group authentication is to authenticate $n$ users at the same time. 
The communication complexity of such an authentication process is expected to be $O(1)$.
\\
\indent In the next section, several related works are mentioned. In general, the researchers' objective is to find a way to authenticate users who belong to the same group at the same time. But current mobility of the users is extremely high and it will be more in the near future. Therefore; one user who belongs to a group will travel to the area of other groups and will try to establish communication with other groups.
Hand-over of users between different authentication groups is still a dilemma for group authentication studies. 
\\
\indent One of the most important problems in the group authentication methods is that the members of the group share their secret key with each other as plaintext messages. The same problem applies to hand-over methods. Any attacker can use plaintext messages to execute various attacks or obtain secret keys. Also attackers can use these plaintext messages to be included in the hand-over process. The proposed method hides the secret during the communication while employing ECC and bilinear mapping. Our study provides solutions for most of the gaps in the literature. One of them is that the researchers who studied group authentication did not take into consideration of hand-over of members between groups. We propose both an authentication method in a group, a key aggrement protocol and a hand-over method. The advantages of our hand-over solution are the low computational cost and its distributed structure. A node should only compute one elliptic curve multiplication in order to access the new group. And also in the new group, any member can perform the hand-over process of new member. There is no need for a central authority for the hand-over process. 
\\
\indent Another gap in the literature is that node compromise, replay, denial of service (DOS) and man-in-the-middle attacks which are the vulnerabilities of wireless networks and there is no proposed solution to overcome these problems. Many group authentication schemes are also vulnerable to the man-in-the-middle attack. The attacker may interrupt the communication of two members within the group and can capture some credentials in order to participate authentication process. Our proposal provides a solution for man-in-the-middle attacks by using bilinear mapping, as discussed in the security analysis section of the study. Overall, our proposed approach for group authentication includes efficient hand-over process, resistant to replay and man-in-the-middle attacks, low computational cost, authentication for mobile and distributed groups.
\\
\indent This paper is organized as follows. The following  section provides an overview of related works about group authentication and hand-over. In the third section, the proposal method for group authentication and authentication between two nodes from different group is presented. The security analysis of our proposal is given in Section IV and the performance evaluation is provided in the following section. The study is completed by conclusion and future works.

\section{Related Work}
Authors propose a group authentication and key agreement protocol for LTE networks in \cite{2}. Only one mobile end device can be authenticated by the serving network. Therefore; the protocol is one-to-one authentication process and it is not applicable for densely populated distributed networks due to the time and resource limitations. Moreover, when one mobile node wants to communicate with other group, the authentication process should be repeated for the other group.
\\
\indent Another group authentication scheme is proposed in \cite{3}. They use a hash function with a pre-shared key (HMAC) in order to authenticate nodes. At the second phase of the authentication process, each user sends a reply to the authentication point at different times.
The second process makes the protocol one-to-one authentication scheme.
\\
\indent A novel method on handover problem for wireless networks is proposed in \cite{4}.  In the architecture, authentication and authorization server shares the secret both with base stations (BS) and mobile stations (MS). In the study, each MS needs to repeat the authentication process with the BS to have a group authentication. But this kind of authentication takes too much time and resource for distributed networks. Also, there is no proposal for authentication between MSs connected to the different BSs.
\\
\indent HashHand \cite{5} is another proposal to hand over nodes between access points in mobile networks. The proposal is a good example of implementation of ECC and bilinear mapping for hand-over purposes. Mobile nodes only consumes source in order to calculate bilinear pairing for authentication code. The most source consuming jobs are done by the authentication server and the structure is not group-based. Therefore; we can assume the proposal a centralised authentication method.
\\
\indent ECC with RSA algorithm is used in \cite{6} in order to overcome with the vulnerabilities in HashHand. The algorithm works faster than HashHand and uses less computational power. But it is still a centralised authentication method. 
\\
\indent Another hand-over method in centrally managed systems is the PairHand  method \cite{7}.   When a mobile node wants to connect with another access point, it calculates a value using its private key and the new access point's public key and shares it with the access point. The access point confirms the value with its private key and the public key of the sending mobile node. The method is not a group-based authentication solution.
\\
\indent The same authors show that PairHand's solution is vulnerable to session key compromise attack in the same year \cite{8}. They produce a solution to the problem of Pairhand algorithm. They recommend that the mobile node in the Pairhand algorithm should send a timestamp before starting the authentication process with the access point. 
\\
\indent Conference key distribution system (CKDS) is proposed in order to create a secret between $n$ members in a group \cite{9}. However, this method is one-to-one rather than a many-to-many method and causes a huge amount of time and resource consumption.
\\
\indent Authors propose a method in order to integrate control and non-payload communication link which is used between unmanned aerial vehicle (UAV) and ground control station (GCS) into LTE network in 
\cite{10}. All the credentials are selected and coordinated by an authentication server (AuS) and UAVs have end-to-end connections with server. This proposal is also suitable to authenticate one UAV at once. Therefore; the method is one-to-one authentication \cite{10}.  
\\
\indent The basis of distributed group authentication schemes is that a secret value is divided into pieces and then secret is recovered by using the pieces. The foundation of the studies in this area was built in 1979 by two different researchers. The Shamir secret sharing (SSS) method was proposed by Adi Shamir \cite{11}. In the same year, the concept of key safeguarding was revealed by George Robert Blakley \cite{12}. Both SSS and key safeguarding schemes are called threshold schemes. According to key safeguarding scheme, a secret can be decomposed into shadows and secret can be recovered from any $r$ or more set of the shadows. But no one can have any information about secret by having $s$ or fewer set of the shadows ($r=s+1$) \cite{13}.
\\
\indent Asmuth and Bloom propose a key safeguarding scheme, which is based on the Chinese remainder theorem (CRT). If anyone has shadows upto $r$, $y$ can be computed easily using CRT and then secret can be recovered. But anyone who has $r-1$ shadows can not recover the secret \cite{13}.
\\
\indent Another secret sharing method \cite{14} is developed using Gray code and XOR operations. The recommended method is for a group of 7 users. 3 or 7 of these 7 group members should come together in order to recover the master key. Although it is seen as a secure method, it is not stated how to share the secret key securely between these members. By eavesdropping to these communications, any attacker can capture secret keys and calculate the master key. At the same time, there is no solution for more than 7 participants.
\\
\indent Harn proposes an algorithm for group authentication in \cite{15}. The algorithm is built based on the SSS. The authentication is not one-to-one type authentication as currently used authentication methods. The algorithm provides authentication for several nodes at the same time. This is called many-to-many authentication type. One of the nodes selects a random polynomical $f(x)$ of degree $t-1:f(x)=a_{0}+a_{1}x+...+a_{t-1}x^{t-1} \mod p$ where $p$ is a prime number. The secret for the communication is $a_{0}$ which is the constant term of the polynomial. The node calculates one secret and one private key for each nodes in the group. Then, the node distributes the keys to the nodes in the group. Each group calculates the secret by lagrange interpolating formula. In the algorithm, many-to-many authentication is done. However; there is no proposal for hand-over of nodes between two different groups.
\\
\indent The authors propose an algorithm by using Paillier threshold cryptography in \cite{16}. 
They compare their result with Harn group authentication method and present the results from their experiments. The results from \cite{16} show that their algorithm has a better computational time than the Harn group authentication algorithm. But they don't take into account the computational cost of public and private key encryptions.
 They also don't propose any method for hand-over of nodes between two different groups. 
\\
\indent Paillier threshold cryptography method is used in \cite{17} in order to authenticate many devices at once. It is not specified in the article how to distribute private keys securely. 
\\
\indent Chien \cite{18} shows that the Harn schemes allow some attacks. If an attacker can get $k$ distinct values in $k$ different trials, the secret function chosen by group manager (GM) can be solved and all users' secret  can be obtained. Chien proposes a new method based on SSS, ECC and pairing-based cryptography in order to ensure a secure group authentication process. According to proposal, GM selects two additive group $G_1$, $G_2$ and one multiplicative group $G_3$ with order $q$. GM makes a generator $P$ for $G_2$ public. A polynomial with degree $t-1$ is chosen. The constant term of the polynomial will be the master secret $s$. The value of $$Q=s\cdot P$$ is computed and shared publicly. For each user, one public key $x_i$ and one private key $f(x_i)$ are chosen and shared with related users secretly. Users participating the authentication phase agree on a random point $R_v$ on $G_1$ in authentication phase. Then, each user computes $c_i=f(x_i)\prod^{m}_{r=1, r\neq i}\dfrac{-x_r}{x_i-x_r}$ and releases $c_i$ $R_v$. After all users release the $c_i\cdot$ $R_v$, each user computes $$\sum^{m}_{i=1}c_i\cdot R_v$$ and verifies if $$e\left(\sum^{m}_{i=1} c_i\cdot R_v,P\right){\stackrel{?}{=}}  e(R_v,Q) $$ holds. The algorithm provides security for group authentication except node compromise and DOS attack. On the other hand it is resource consuming method for users. Chien also don't propose any hand-over algorithm in his study as well.

\section{Proposed Method}
\indent In our proposal, we use the same $(t,m,n)$ logic as in Harn's algorithm. There are $n$ users in the group and $m$ users want to authenticate each other. $t$ is the threshold for the algorithm ($t<m<n$). $n$ should be greater than $m$ and the secret can be obtained by the participation of $m$ or more users. 
\\
\indent
It should be noted at this stage that the proposed method can especially be in use for a public safety networks (PSN) and the Internet of Things (IoT) networks. More than one group takes part in our scenario. Each group has a group manager denoted by GM. GM is assumed to be infrastructure-based and does have relatively more computational power. All group managers can communicate with each other securely via traditional cryptographic methods. In addition to the group managers, each group has several other members which have resource or computational constraints.
\\
\indent Note that if the PSN environment is under consideration, GM is basically the ground radio stations (GRS) and the group members are UAV devices. Similarly, gateways with specific capabilities in an IoT environment are GMs and radio frequency identification tags can be considered to be other  members in a group. The capabilities of tags and UAV devices are at a certain restricted rate. Under these considerations we propose a novel method. The proposed method has three stages. The first stage involves authentication which is based on ECC and SSS. This first stage consists of two phases, which are called the initialisation and the confirmation phases. The second stage, which is the key agreement stage, provides a solution to construct a master key for further communications. And the hand-over stage is a crucial part of group communication in order to authenticate the users from other groups. The details of each phase are presented at below.
\begin{center}
  {\bf The Initialisation Phase:}
\end{center}

\begin{enumerate}
	\item GM selects a cyclic group $G$ and a generator $P$ for $G$.
	\item GM selects a bilinear map $e: G \times G\rightarrow G'$ and an $E=Encryption()$ and $D= Decryption()$ algorithms.
	\item A polynomial with degree $t-1$ is chosen by GM and the constant term is determined as master key $s$.
	\item GM selects one public key $x_i$ and one private key $f(x_i)$ for each user in the group $U$ where each user is denote by $U_i$ for $i=1,\dots,n$.
	\item GM computes $Q=s\cdot P$.
	\item GM makes $P, Q, e, E, D, H(s)$ public and shares $f(x_i)$ with only user $U_i$ for $i=1,\dots,n$.
\end{enumerate}
\indent \indent The confirmation phase is executed after GM shares the values with the related users. There are two different options in the confirmation phase. One of them is that the GM will be responsible to confirm the group members. In the other case, that is  if GM is not responsible, any member within the group will confirm the other members.\\
\begin{algorithm}
{Each member Compute  $f(x_i)\cdot P$ }
{Share  $f(x_i)\cdot P \|ID_i$ with GM and other members}\newline\\
\If{GM verifies the authentication}
{
	GM computes $f(x_i)\cdot P$ for each user.\newline\\
	\If{All values are valid}
	{
		Print "Authentication is done."
	}
	\Else{
		Repeat.
	}
}
\Else{
	{Any user computes $c_i$=$f(x_i)\cdot P$${\overset{m}{\underset{r=1, r\neq i}{{\displaystyle		
	\prod}}}(-x_r/(x_i-x_r))}$ for each user.}\newline\\
	\If{${\overset{m}{\underset{i=1}{{\displaystyle\sum}}}c_i}$ is equal to Q}
	{
		Print "Authentication is done."
	}
	\Else{
		Repeat
	}
}
\BlankLine
\caption{Confirmation Phase}
\end{algorithm}
\begin{center}
{\bf The Confirmation Phase}
\end{center}
\begin{enumerate}
\item  Each user computes $f(x_i)\cdot P$ and sends  $f(x_i)\cdot P\|ID_i$ to GM and other users ($ID_i$ is the identification number of the user). 
\item If GM verifies the authentication, GM computes $f(x_i)\cdot P$ for each user and verifies whether the values are valid or not.
\item If GM is not included in the verification process, any user in the group computes $$C_i=\left(\prod^{m}_{r=1, r\neq i}\dfrac{-x_r}{x_i-x_r})\right)f(x_i)\cdot P$$ for each user. 
\item  User verifies if $$\sum_{i=1}^{m}C_i  {\stackrel{?}{=}}  Q\text{ holds.}$$
\item If it holds, authentication is done. Otherwise; the process will be repeated from the initialization phase.
\end{enumerate}
\indent \\\indent
Both authentication by GM and any group member is given in the Algorithm 1. It is clear that group members should only compute one elliptic curve multiplication operation. And also users should send their identification numbers by concatenating with public shares in order to avoid confusion for further communications. Because; these public shares will be used by other users in further communications and in the group key agreement stage. All group users should know which public share belongs to which user.  
\\
\indent After authentication is done, users will communicate with each other by using symmetric key encryption. Shared key for symmetric key encryption will be calculated by senders and receivers.
\\
\indent Pairing-based cryptography is used in order to compute shared key between the group members. Bilinear-map is a map which is linear in each component \cite{19}. Let say $P$ and $Q$ is a point on group $G_1$ and $G_2$. If $e(P , Q)$ is equal to $z$, $e(aP , bQ)$ should be  $z^{ab}$. And also $e(aP , bQ)$ is equal to $e(bP , aQ)$.
\\
\indent Let set the key, $K$ as $$K= e((y_iy_j)P , Q)$$ where $y_t=f(x_t)$ i.e., $y_t$ is the secret of the user $U_t$. The sender will use its own private key $(y_i)$ and  the value sent 
by receiver $(y_jP)$ and the public information $Q$. The receiver will obtain the same key by using its own private key $y_j$, value sent by sender $(y_iP)$ and $Q$.
\\
\indent After this stage, group members can communicate with each other by a symmetric key encryption method. But instead of using different keys for each user, the master key that was selected by GM can be used as the group key. The problem is how the users will recover the master key. We basicly exploit SSS and a symmetric key encryption method to share the master key in the group key agreement stage.\\
\begin{algorithm}
{$U_i$ computes $E_{e(f(x_i)f(x_j)P),Q)}[f(x_i)]$ for each $U_j$.}\newline\\
{Each user computes $D_{e(f(x_j),f(x_i)P),Q)}[f(x_i)]$.}\newline\\
{Each user computes $$s'=(\sum_{i=1}^{m}f(x_i)\prod^{m}_{r=1, r\neq i}\dfrac{-x_r}{x_i-x_r})$$}\newline\\
{Each user computes $H(s')$.}\newline\\
\If{$H(s')$ is equal to $H(s)$}
{
	Print "Master Key is recovered".
}
\Else{
	Repeat.
}
\BlankLine 
\caption{The Group Key Agreement Stage}
\end{algorithm}
\\\\
\begin{center}
{\bf The Group Key Agreement Stage}
\end{center}
\begin{enumerate}
\item Each user shares its own secret key $f(x_i)$ with other users using symmetric key encryption.
\item Each user decrypts the values and obtains $m$ different $f(x_i)$.
\item Each user computes $$s'=\sum_{i=1}^{m}f(x_i)\prod^{m}_{r=1, r\neq i}\dfrac{-x_r}{x_i-x_r}$$
\item Each user verifies $$ H(s')  {\stackrel{?}{=}}  H(s)\text{ holds.}$$ 
\end{enumerate}
\indent \\\indent At the end of the group key agreement stage each member within group will recover the master key as given in the Algorithm 2.
\indent After the group key agreement process, the members of the group will be able to communicate with each other  using master key. In addition GM can update $x_i$ and $f(x_i)$ values remotely using master key in order to avoid the replay attacks mentioned in  the security analysis part of the study.  
\\
\indent GM always knows that $m$ user participated the authentication and $x_m$ values were used so far. If GMs can coordinate the $x$ values which they used for group authentication, they will use distinct $x$ values for each user. If GMs select different $x$ values and share their polynomial with other GMs, the hand-over process can be done as given in Algorithm 3. 
\\
\indent In many studies, group authentication was completed at this point. However, since UAV and IoT nodes are constantly on the move, they will be able to access the coverage area of another group or the IoT gateway. Instead of repeating the entire process, it is necessary to quickly authenticate the new member. Therefore, each group authentication scheme should have a hand-over method.
\begin{center}
{\bf The Hand-Over Stage}
\end{center}
\begin{enumerate}
\item $GM_1$ shares group-1 polynomial $f(x)$ with $GM_2$ by secure channel.
\item $GM_2$ shares group-2 polynomial $g(x)$ with $GM_1$ by secure channel. 
\item If $GM_2$ is responsible for hand-over, the user $U_i$ ,which wants to participate Group-2, computes $f(x_i)P_2$ and shares $x_i$, $f(x_i)P_2$ with $GM_2$ ($P_2$ is public).
\item $GM_2$ verifies $f(x_i)P_2$ is correct.
\item If it is correct, $GM_2$ shares the encryption of Group-2 master key ($E_{e(s_2f(x_i)P_2,Q_2)}[s_2]$) with $U_i$. 
\item $U_i$ computes  $D_{e(f(x_i)s_2P_2,Q_2)}[s_2]$ and gets master key of Group-2 for further communications ($P_2$ and $Q_2$ are public).
\item If $GM_2$ is not responsible for hand-over, $U_i$ requests $g(x_i)$ from $GM_1$. 
\item $GM_1$ computes $g(x_i)$ and share with  $U_i$ securely.
\item $U_i$ computes $g(x_i)P_2$.
\item $U_i$ shares $x_i$ and $g(x_i)\cdot P_2$ with any user of Group-2 ($U_j$).
\item $U_j$ computes $$Q_2'=(\sum_{i=1}^{m+1}g(x_i)P_2\prod^{m+1}_{r=1, r\neq i}\dfrac{-x_r}{x_i-x_r})$$\newline\\
\item $U_j$ verifies $$ Q_2'  {\stackrel{?}{=}}  Q_2$$ holds.
\item If it holds, $U_j$ shares its public key $g(x_j) P_2$ and the encryption of group-2 master key ($E_{e(g(x_j)g(x_i)P_2,Q_2)}[s_2]$) with $U_i$.
\item $U_i$ computes  $D_{e(g(x_i)g(x_j)P_2,Q_2)}[s_2]$ and gets master key of group-2 for further communications.
\end{enumerate}

\begin{algorithm}
\If {$GM_2$ is responsible}{
	$U_i$ shares $f(x_i)P_2$ with $GM_2$.\newline\\
	$GM_2$ verifies $f(x_i)P_2$ is correct.\newline\\
	\If {The value is correct}{
		 $GM_2$ shares $s_2$. \newline\\
	}
	\Else
	{
		Print "Not valid user."
	}
}
\Else
{
	{$GM_1$ computes $g(x_i)$ and share with  $U_i$ }\newline\\
	{$U_i$ computes $g(x_i)P_2$}\newline\\
	{$U_i$ shares $x_i$ and $g(x_i)P_2$ with any user of group-2 ($U_j$).}\newline\\
	{$U_j$ computes $$s_2'=(\sum_{i=1}^{m+1}g(x_i)\prod^{m+1}_{r=1, r\neq i}		\dfrac{-x_r}{x_i-x_r})$$ } \newline\\
	{$U_j$ computes $H(s_2')$.}\newline\\
	\If{$H(s_2')$ is equal to $H(s_2)$}
	{
		Print "Valid user".\newline\\
		$U_j$ shares $s_2$.\newline\\
	}
	\Else
	{
		Print "Not valid user."
	}
}
\BlankLine 
\caption{Hand-Over Stage}
\end{algorithm}
\indent \\\indent Overall, we propose a comprehensive solution for authentication of users belong both to the same group and to the different groups in three different stages. A group authentication is accomplished with very low computational power on users in the first stage. A master key is recovered by all group users for a distributed environment in the second stage. In the last stage, a user is authenticated by the new group in very short time period. The details of the security and performance analysis is given in the next sections of the study. 
\\
\section{Security Analysis}
In this session, we analyze certain possible attacks to the presented algorithms above.\\
\textbf{Theorem 1:} \textit{Group authentication cannot be performed without t valid public and private values.}
\\
\textit{Proof.} Since the stated polynomial $f(x)$ is of degree $t-1$, it is necessary to know $t$ distinct pairs   of ($x$,$f(x)$) for the formation of the polynomial again. Polynomial cannot be formed again by holding less than $t$ pairs.
\\
\textbf{Theorem 2:} \textit{The attacker who capture the value of $Q$ and $P$  sent by the group manager publicly cannot have knowledge of secret $s$.}
\\
\textit{Proof.} Given two points $P$ and $Q$ on an elliptic curve group, it is hard to find the $s$ value that provides a relationship like $Q = s\cdot P$. This open problem is called Eliptic Curve Discrete Logarithm Problem (ECDLP). Therefore, it hard to find $s$ by having $Q$ and $P$.
\\
\textbf{Theorem 3:} \textit{The attacker who capture the value of $f(x_i)P$  sent by the group members to the group manager cannot have knowledge of $f(x_i)$.}
\\
\textit{Proof.} Due to the hardness assumption of ECDLP, it is hard to find $f(x_i)$ by having $f(x_i)P$.
\\
\textbf{Theorem 4:} \textit{The attacker can capture  $f(x_i)P$, e and $Q$ but can not obtain a valid symmetric key in order to establish a communication with user $U_i$.}
\\
\textit{Proof.} The attacker will need $f(x_j$)  to compute $e(y_i.y_j.P,Q)$ but $f(x_j)$ is a secret known only by the user  $U_j$.
\\
\textbf{Theorem 5:} \textit{The attacker can perform man-in-the-middle attack but can not have any credentials.}
\\
\textit{Proof.} Attacker can intercept the communication between two users and act as a real user. The attacker can continue to participate the process till the bilinear mapping phase. Because the attacker have only $f(x_j) P$, $f(x_i)P$ and $Q$, the key $(e(f(x_i)f(x_j)P,Q))$ that is used for the construction of master key can not be obtained.
\\
\textbf{Vulnerability 1:} \textit{If the authentication secrets are used more than one times, attacker can perform replay attack in the next trail.}
\\
\textit{Proof.} The attacker can eavesdrop the traffic in the first trail and capture $f(x_i)P$. In the next trail the attacker can send $f(x_i)P$ to GM before $U_i$ and involve to the group. In order to avoid this vulnerability, GMs should update credentials using master key for each group authentication.
\\
\textbf{Vulnerability 2:} \textit{The attacker can perform DOS attack for authentication process.}
\\
\textit{Proof.} Attacker can share a not-valid value when the members send their shares to user which will control the authentication. User can not compute a valid value and repeat the process. Attacker can share not-valid value again and perform denial of authentication.\\
\textbf{Vulnerability 3:} \textit{The node compromise attack can be performed.}
\\\\
\textit{Proof.} If the attacker could physically capture a group member, it obtains the secret key of the member. As a result of the capture of the secret key, the attacker can generate a valid public key and share it with GM in order to authenticate itself. If it has a secret key, it also can communicate with the other members of the group by producing symmetric keys.
\\
\begin{figure}
\centering
\begin{tikzpicture}
\begin{axis}[xlabel={The Number of Users}, ylabel={Computational Cost $T_{mul,q}$}
,axis lines=middle
,samples=41, grid, thick
,domain=100:300
,legend style={at={(0.5,-0.2)}, anchor=north,legend columns=-1}
,xtick={0,20,...,300}
,ytick={0,1000,3000,6000,8000,10000}
]
\addplot+[mark=*,mark repeat=2]
{x*7+6785};
\addlegendentry{Chien[16]}
\addplot+[mark=*,mark repeat=2]
{x*45+1418};
\addlegendentry{Harn[13]}
\addplot+[mark=*,mark repeat=2]
{x*0+1189};
\addplot+[mark=nomark]
{x*0+0};
\addlegendentry{Proposed Method}
\end{axis}
\end{tikzpicture}
\caption{The Comparison of Computational Costs in Authentication Stage} \label{fig:M1}
\end{figure}
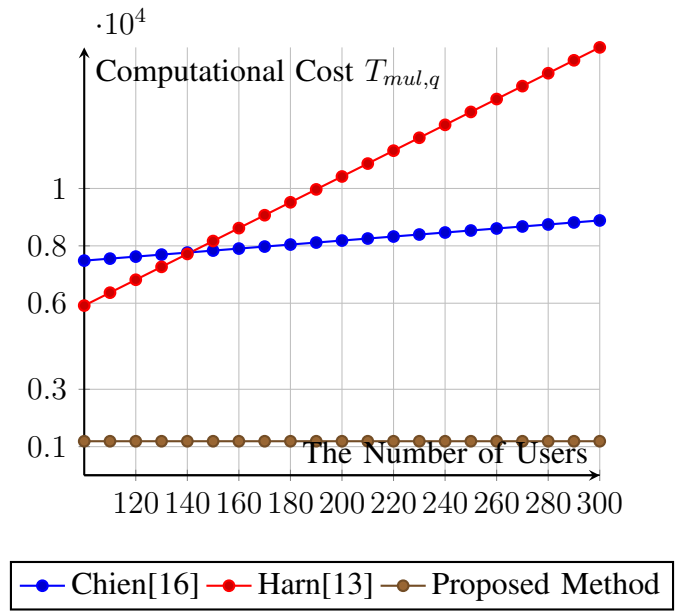
\textbf{Vulnerability 4:} \textit{The group members can perform DOS attack for confirmation point.}
\\
\textit{Proof.} If the group members send their shares with the point which is responsible for confirmation at the same time, the confirmation point can be locked. The solution for this kind of DOS attack is still a challenge in group authetication studies. 
\section{Performance Analysis}
We  use the time complexity approach in \cite{18} to compare our algorithm with Harn and Chien schemes. Both our algorithm and Harn or Chien algorithms have a GM which is responsible for initialise the authentication. In a PSN or wireless sensor network, the group manager will be a GRS and group members will be UAVs or sensors. Therefore; group members will have computational and resourse restrictions. 
\\
\indent Due to the reasons we mentioned before, we only take into consideration the computations that are made by group members. While each user in Chien algorithm should compute (7m+6785)$T_{mul,q}$ \cite{18}, each user in Harn asynchronous multiple authentication scheme should compute (45m+1418) $T_{mul,q}$ [18]. ($T_{mul,q}$ denote the time for one multiplication in field $q$ where $q$ is 160 bits, m denote the number of user in the group.)    
\\
\indent In our proposal the group members should only compute one elliptic curve point multiplication  ($T_{EM}$). According to Chien \cite{18}, 1 $T_{EM}$ is roughly equal to 29  $T_{mul,p}$ ($T_{mul,p}$ denote the time for one multiplication in field $p$ where $p$ is 1024 bits). The security of ECC with 160-bit key is roughly equivalent to that of RSA with 1024-bit key or D-H algorithm with 1024-bit key. Therefore; 1 $T_{mul,p}$ is roughly equal to 41 $T_{mul,q}$ \cite{18}. In our authentication algorithm, group members compute 29 $T_{mul,p}$, which is 1189 (29x41) $T_{mul,q}$.
\\
\indent Confirmation for authentication process is done by group members in Chien and Harn schemes. But in our scheme, the GM or only one user is responsible for the confirmation part of the authentication. As you can see from the Fig. 1, our proposal is scalable with the number of group members.
\\
\section{Conclusion}
The study proposes a novel method for authentication and hand-over process on group communication in wireless networks. Many-to-many authentication is used for group authentication by several studies but resource-constrained users were forced to compute more than their capacity. 
Group members should only compute one elliptic curve point multiplication in the proposed method. Most of the resource-consuming work is done by the GM or one of the group members not all the group members as other proposed methods. 
\\
\indent The vulnerabilities which we mentioned in security analysis part are still research area for scientists who study on secret sharing algorithms in group communication. As far as we know there is no proposal for replay, node compromise and DOS attacks under the framework of secret sharing schemes. Our proposal provides the security for replay attacks if the GMs update the credentials for each authentication. 
\\
\indent Our study is made by assuming that the group manager or base station is infrastructure based. For this reason, there is no computation or resource restriction of the base station. However; the base stations are gradually getting mobile and infrastructureless. New methods  are needed  to deal with these challenges.
\\
\indent SSS and ECC are used on the basis of the proposed algorithms. ECC method is more cost effective than other public key cryptography methods. ECC can be used to perform by the devices with resource and computational restrictions. But even this single operation creates a certain load on the devices. One future work is to find a cross-layer solution that will allow users to send their private keys secretly.




\end{document}